\documentclass[journal=jctcce,manuscript=article]{achemso} 
\setkeys{acs}{articletitle = true} 

\usepackage[version=3]{mhchem}   
\usepackage[hidelinks]{hyperref}
 
\author{Maria Ley-Flores}
\affiliation{Pritzker School of Molecular Engineering, University of
Chicago}
\author{Archit Chabbi}
\affiliation{Department of Bioengineering, Rice University}
\author{Riccardo Alessandri}
\affiliation{Pritzker School of Molecular Engineering, University of
Chicago}
\author{Sam Marsden}
\affiliation{Pritzker School of Molecular Engineering, University of
Chicago}
\author{Isabella Vettese}
\affiliation{Department of Chemistry, University of
Chicago}
\author{Stuart J. Rowan}
\affiliation{Department of Chemistry, University of
Chicago}
\alsoaffiliation{Pritzker School of Molecular Engineering, University of
Chicago}
\author{Juan J. de Pablo}
\affiliation{Pritzker School of Molecular Engineering, University of
Chicago}
\alsoaffiliation{Materials Science Division, Argonne National Lab}

\email{mleyf@uchicago.edu}

\title[]{Numerical Study of Cleavable Bond-Modified
Polyethylene for Circular Polymer
Design}

\begin{document}

\begin{abstract}
There is considerable interest in designing new polymeric materials with built-in mechanisms for recycling. In this work, we present a systematic exploration of several polyethylene-based polymers that contain strategically incorporated cleavable bonds that are susceptible to chemical degradation. We consider ten distinct telechelic functionalities across varying chain lengths: ester, aromatic ester, anhydride, carbonate, urethane, siloxane, oxalate, acetal, urea, and amide linkages. Our study is focused on the effects of cleavable bonds on density, diffusion, enthalpy, and melting temperatures. We also examine the crystallization process. These are key properties for polymer processing and, by comparing them to those of linear polyethylene, we establish a platform that will help guide future synthetic efforts towards the design of sustainable plastics with performance comparable to that of traditional, high-performance polyolefin-based materials.
\end{abstract}


\section{Introduction}
Plastic waste management has emerged as a pressing global challenge. 
Polyolefins constitute the largest fraction of global plastics production.\cite{giacovelli2018single, geyer2017production, jambeck2015plastic} The majority of current plastics recycling efforts rely on mechanical processes, often leading to performance losses and short usage cycles.\cite{ragaert2017mechanical} Chemical recycling technologies have been developed \cite{vollmer2020beyond, lopez2017thermochemical, epps2021sustainability, hinton2022innovations, zhao2022plastic} to break down polymers into small, more versatile molecules that can be used for a wide range of potential applications, including as feedstocks for new polymers.\cite{coates2020chemical} However, due to the high energetic cost associated with breaking down the carbon-carbon $\sigma$ bond, progress in chemical recycling of polyolefins has been particularly challenging. Promising new approaches have been developed for enzymatic based degradation of plastics but, to date, such processes have only been demonstrated for polyethylene terephthalate (PET)-based polymers.\cite{tournier2020engineered} 

One approach to address the challenge of polyolefin recycling is to introduce cleavable functional groups within the structure of linear polyethylene (PE).\cite{coates2020chemical,stempfle2016long, haussler2021closed} Such long-spaced aliphatic polycondensates, also termed "polyolefin-like materials," offer the potential to open the door to polymer circularity while displaying properties similar to those of commodity plastics. In a previous review paper, Stempfle \emph{et al}\cite{stempfle2016long} provide a comprehensive account of the physical properties of the long-spaced aliphatic condensates that have been studied to date. Some trends have been observed; for example the melting point increases with increasing hydrocarbon chain length between cleavable bonds in some of the long-spaced aliphatic polycondensates considered in the literature. There are, however, disparities in the reported values among various authors, which can probably be attributed to differences in synthetic approaches and sample preparation methods. Importantly, key properties such as the density or transport properties of the materials, which are critically important to their performance in plastic processing equipment, have not been examined before in a systematic manner.

Past work has established that molecular models provide a valuable tool to bridge the gap between limited experimental data and property prediction for PE-like materials. Harmandaris \emph{et al},\cite{harmandaris2003crossover} for example, have computed the self-diffusion coefficient of polyethylene melts to determine the crossover from the Rouse to the entangled melt regimes. Mondello \emph{et al}\cite{mondello1998dynamics} performed non-equilibrium molecular dynamics (NEMD) simulations to calculate the Newtonian viscosity of higher n-alkanes at 448 K. Ramos \emph{et al}\cite{ramos2008entanglement} combined computer simulations and experiments to describe the viscoelastic properties of linear PE. However, although PE is undoubtedly the most studied polymer in the polymer physics literature,\cite{ramos2018predicting} there is no precedent for computational studies of long-spaced aliphatic polycondensates.

In this work we present a computational study of the effect of cleavable bonds in ten different linear polycondensates. We consider anhydride, amide, urethane, urea, acetal, carbonate, ester, siloxane, and oxalate functionalities. These functional groups are significantly high in energy, and they lower the energy barriers for degradation via solvolysis. By using MD simulations, we study the effect of chain length on the density, heat of vaporization, and diffusivity in the melt. We also simulate the crystallization process of the resulting semi-crystalline samples of the PE-like candidates considered here. The properties of these polycondensates are compared to those of linear polyethylene at different temperatures, thereby providing a basis to evaluate their potential as sustainable alternatives in the context of polymer processing.

\section{Methods}
\subsection*{Molecular Models}

\begin{figure}[h!]
    \centering
    \includegraphics[width=5in]{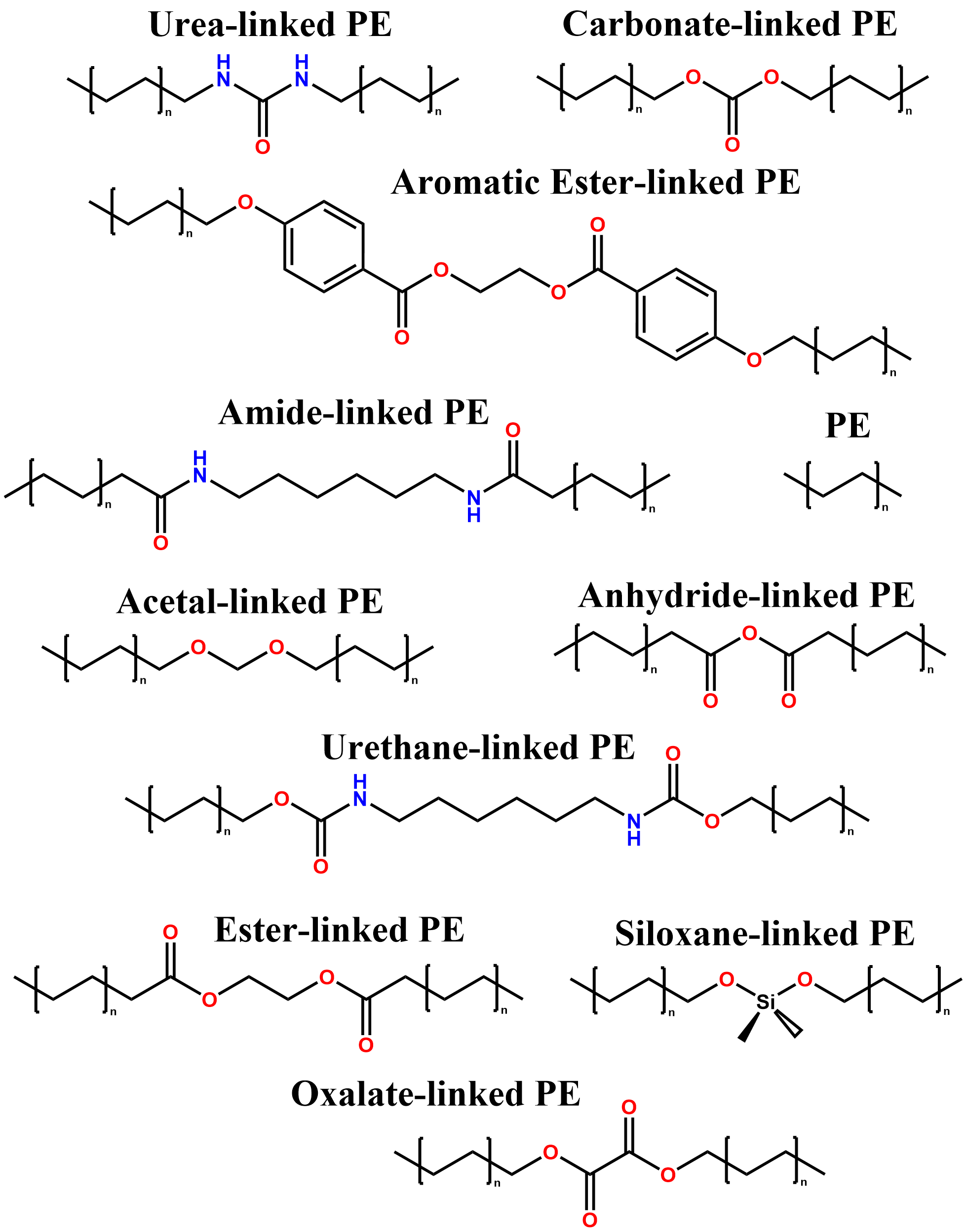}
    \caption{Polymer chains studied in this work: polyethylene and cleavable bond-modified polyethylenes featuring ten different functionalities at the center of the backbone.}
    \label{fig:structures}
\end{figure}

In order to study a wide range of chain lengths, a united
atom (UA) model was used to simulate the PE and PE-like chains. In
this model, the hydrogen atoms are not explicitly simulated. Rather, they are grouped with the carbon to which they are
bonded into a single interaction site. Previous studies have shown that united-atom models are better than all-atom models at reproducing thermodynamic and transport properties for long hydrocarbons.\cite{da2020all,da2022all} 
In particular, we used the GROMOS 54A7\cite{schuler2001improved} force field. All molecules were parametrized using semiempirical QM calculations from the MOPAC software \cite{dewar1977ground, stewart2013optimization} implemented by the Automated Topology Builder (ATB)\cite{stroet2018automated} server tool.
Models will be made available via the polyply~\cite{grunewald2022polyply} library.
We modeled chains containing an individual cleavable bond, surrounded by methylene segments from 20 to 80 united atoms. A total of 47 different chains were considered (4 chain lengths for 10 functionalities, and 7 chain lengths for PE). To facilitate comparison to furture expermental work, each of the PE-like chains were modeled as symmetric molecules with the cleavable bond placed in the middle of the chain (Figure \ref{fig:structures}).

\subsection*{Simulation Methods}

We seek to describe the temperature and molecular weight dependence of the density, heat of vaporization, and self-diffusion coefficient for PE-like materials featuring different functionalities proposed in the literature. MD simulations were performed with the GROMACS \cite{van2005gromacs, pronk2013gromacs, berendsen1995gromacs, abraham2015gromacs} software version 2022.4. The equations of motion were integrated using the velocity Verlet algorithm with a 5 fs time step.

The simulation box was generated by randomly inserting between 100 and 200 chains of a particular polymer in a cubic cell of 40 nm on a side. We used a box with 100 chains for the longest chain and 200 chains for the smallest molecules; these systems are sufficiently large to avoid finite-size effects. After an initial energy minimization at 298 K to remove high-energy configurations of the initial structures, the system was rapidly condensed at 523 K and 500 bar in the isothermal-isobaric (NPT) ensemble for 500 ps. The last configuration was then used to equilibrate individual systems for 10 ns at the corresponding pressure and temperature condition using the NPT ensemble. The last 5 ns were used to compute the average density of the liquid polymer melt.

In general, barostats, which manipulate positions through volume changes, can interfere with the dynamics of the system, leading to biased transport properties.\cite{maginn2019best} Basconi and Shirts\cite{basconi2013effects} revealed that the estimates of self-diffusivity in the microcanonical (NVE) ensemble are statistically indistinguishable from those obtained by canonical (NVT) simulations with the velocity-scaling thermostats employed in this work. Therefore, the final configurations of the NPT simulations were used as starting configurations for a 100 ns NVT run to compute the self-diffusion coefficient of each system by means of the Einstein relation (Equation \ref{eq:Einstein-diff}). Altogether, the simulations spanned a temperature range from 373 K to 573 K. Whenever a temperature control was necessary, the velocity-rescaling thermostat \cite{bussi2007canonical} was employed. For controlling the pressure, the Parrinello-Rahman barostat \cite{nose1983constant, parrinello1981polymorphic} was used. 

\begin{equation}
\label{eq:Einstein-diff}
\lim_{t\to\infty} \langle \|\mathbf{r_i}(t) - \mathbf{r_i}_0 \|^2\rangle_i= 6Dt
\end{equation}

Additionally, three single-molecule simulations of each PE-like chain at 150 K, 373 K, 473 K, and 573 K were performed for 100 ns with an NVT ensemble to calculate the average potential energy of the gas phase. This value was then combined with the potential energy of the liquid simulations to estimate the heat of vaporization by means of Equation \ref{eq:HeatVap}:

\begin{equation}
\label{eq:HeatVap}
\Delta H_{vap}= E^{single}_{pot} - \frac{E^{bulk}_{pot}}{N_{mol}} + RT
\end{equation}

\begin{figure}[ht]
    \centering
    \includegraphics[width=0.5\textwidth]{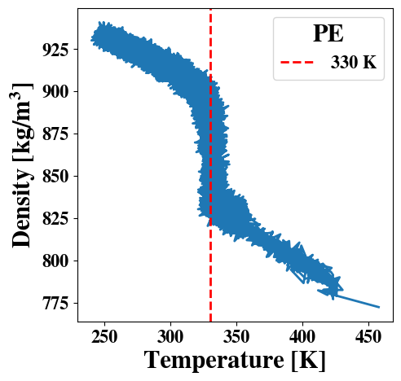}
    \caption{Crystallization temperature of the polymers, determined by estimating the intersection of two lines fitting the melt and the semi-crystalline regions of the material from a plot of density vs temperature.}
    \label{fig:crystal_protocol}
\end{figure}

To examine the crystallization processes of the polymer chains, a cooling process was performed from an initial amorphous melt at 423 K to a final temperature of 150 K using a piece-wise linear function at a cooling rate of 0.05 K/ns. All annealing simulations were performed for the longest chains consisting of a cleavable bond and 80 methylene units. The crystallization temperature was determined by estimating the intersection point of the two fitted-lines from plotting the density of the polymer against the temperature (Figure \ref{fig:crystal_protocol}).

\section{Results and discussion}

\subsection*{Density}

\begin{figure}[!hb]
    \centering
    \includegraphics[width=1.00\textwidth]{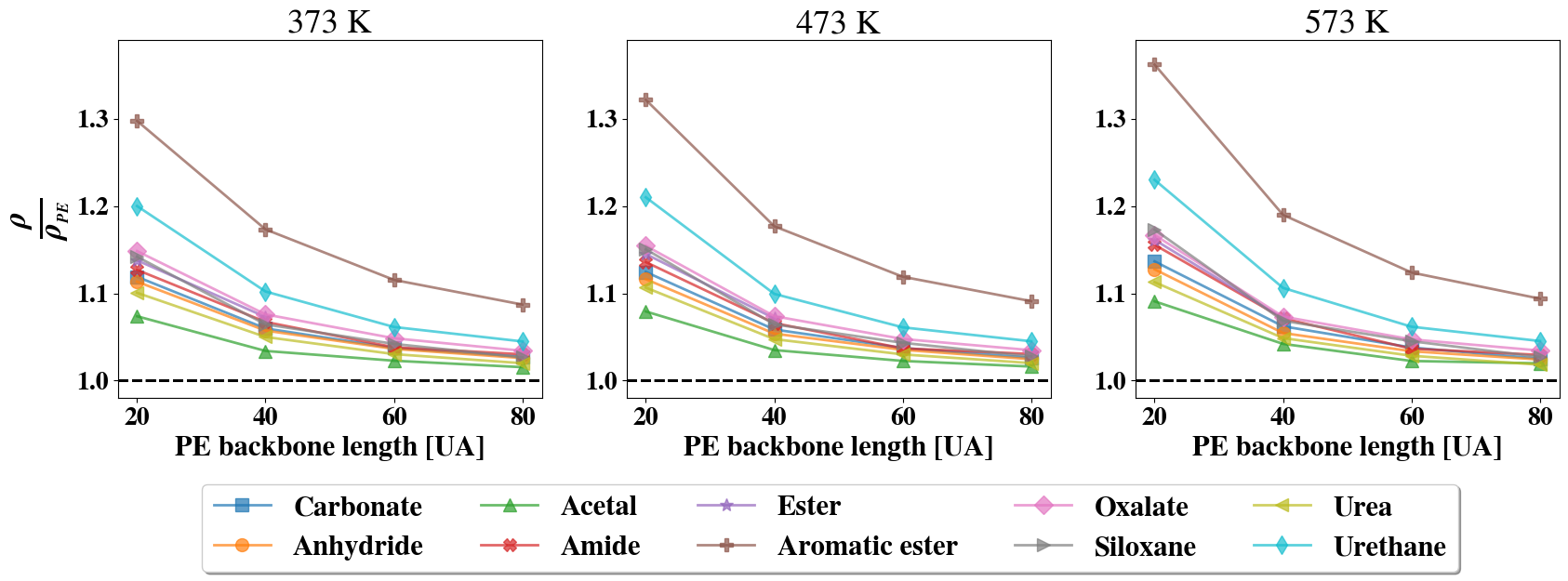}
    \caption{Density of cleavable bond-modified polyethylene normalized by the density of linear polyethylene at 373 K, 473 K, and 573 K.}
    \label{fig:density_gromos}
\end{figure}

Figure \ref{fig:density_gromos} shows the relative density of each modified polyethylene compared to that of linear polyethylene at 373, 473, and 573 K. Overall, by increasing the backbone length, the density converges to a value close to that of linear polyethylene in almost all cases. The intermolecular interactions coming from the single cleavable bond are not strong enough to produce a substantial change in the relative density of modified PE melts. At higher temperatures, however, the differences between the density of different chains become more pronounced. 

The aromatic ester-linked PE shows the most significant increase in density, with deviations of 8.7\%, 9.1\%, and 9.4\% at 373, 473, and 573 K, respectively, for the longest chain length. In contrast, for the ester-linked PE, which also features two ester groups separated by two methylene units, the differences in density are less significant, and similar to the behavior of the chains featuring carbonate, amide, anhydride, urea, urethane, siloxane, and oxalate functionalities. These findings suggest that the introduction of the aromatic ester linkages leads to a more efficient packing of the chains, due to the $\pi$-$\pi$ stacking interactions associated with the two aromatic rings in its structure.

After the aromatic-ester, the urethane linkage exhibits the highest deviations in density with respect to polyethylene. The urethane group displays strong hydrogen bonding interactions, which become more important at shorter chain lengths. At longer backbone lengths, the urethane-linked PE reaches a density that is closer to that of polyethylene, with a deviation of only 4.5\% at the highest temperature. The amide-linked PE, which has a structure similar to that of the urethane, where the functionalities are spaced by six methylene units, displays a less pronounced change in density as a function of backbone length. This is likely due to the higher flexibility that stems from the extra oxygens in the urethane-linked PE, and the higher strength and directionality of the hydrogen bonding in urethane compared to those of the amide groups. 

Acetal functional groups are not sufficiently electronegative to participate in strong van der Waals bonding interactions. As a result, the acetal-linked PE displays the closest behavior to PE. The siloxane linkage, although nearly non-polar, displays a higher density compared to the acetal-linked PE. This is likely due to the differences in molecular weight between the heavier silicon atom and the carbon atom. For the shortest chain length, we can see that the density of the siloxane-linked PE becomes even greater than that of the oxalate-linked PE. Siloxane groups are expected to increase the flexibility of the chain as the Si-O bond length is longer and rotation about Si-O bond is extremely facile.\cite{sun1997polysiloxanes} More flexible chains can adopt a wider variety of conformations that allow them to pack more efficiently and reduce free volume. This is especially true at higher temperatures, where chains have increased mobility.

At lower temperatures, the shortest chains are more densely packed than the reference polyethylene with the same backbone length with differences close to 30\% for the case of the aromatic ester-linked PE. With the exception of the aromatic ester-linked PE, all chains approach the density of polyethylene as the number of backbone atoms increases with differences of less than 5\%. For the aromatic ester-linked PE, the converged density is close to 9\% higher than the predicted value for the reference polyethylene.

\subsection*{Heat of Vaporization}

While the density relates to how tightly the polymer chains are packed, the heat of vaporization provides a measure of the energy required to overcome intermolecular forces that hold molecules together in the melt phase. In general, stronger intermolecular interactions in the form of hydrogen bonding, van der Waals forces, and dipole-dipole interactions result in higher values for the heat of vaporization.

\begin{figure}[!hb]
    \centering
    \includegraphics[width=1.00\textwidth]{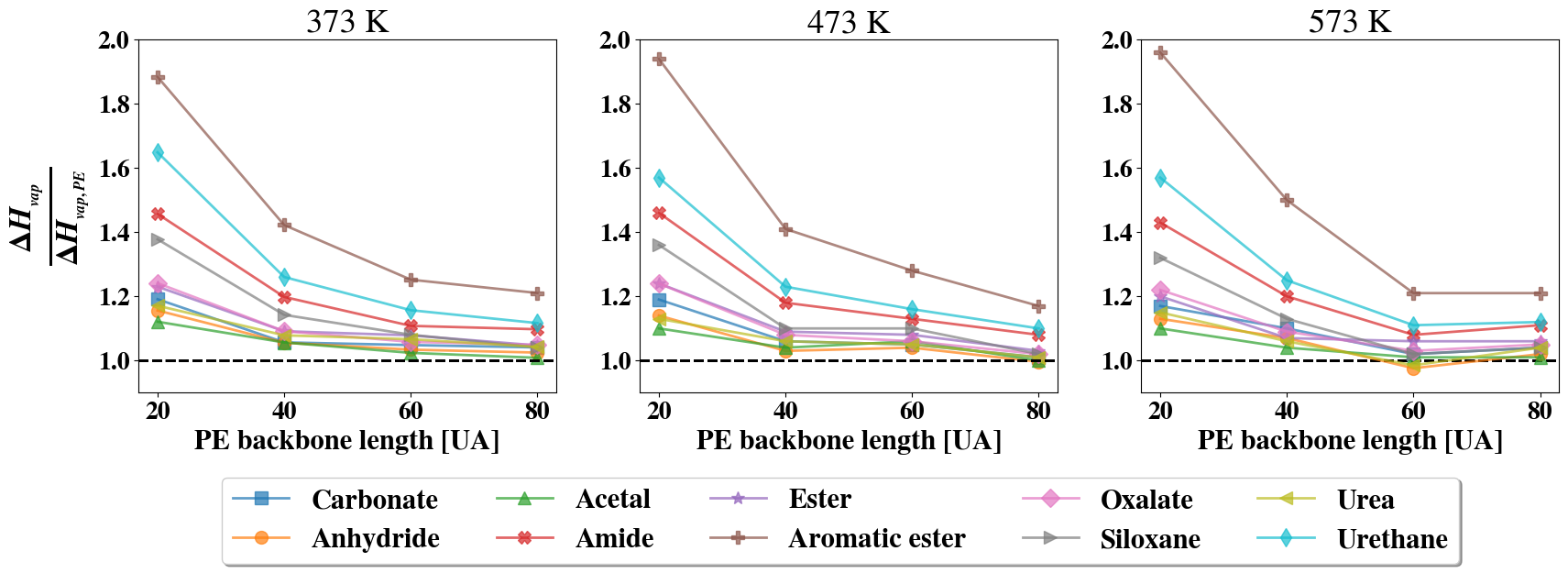}
    \caption{Relative heat of vaporization for cleavable bond-modified polyethylene with respect to linear polyethylene at 373 K, 473 K, 573 K.}
    \label{fig:hvap}
\end{figure}

Figure \ref{fig:hvap} shows the heat of vaporization of the cleavable bond-modified PE relative to PE at 373, 473, an 573 K. Similar to the relative density, the heat of vaporization of almost all modified PE-like molecules converges to a value close to the heat of vaporization of the unmodified PE as the backbone length increases. Likewise, the deviations increase as the temperature increases, leading to values closer to those of polyethylene at lower temperatures and longer backbone lengths.

Just as for the density, the biggest deviations come from the aromatic ester-modified PE. At the shortest chain length, the heat of vaporization of the modified PE is close to 90\% greater than the one of the unmodified PE. As the chain length increases, this deviation reduces to 21\%, compared to the 9.4\% increase in density (Table \ref{tab:heatvap}). This shows how the aromatic ester linkage significantly enhances the strength of the intermolecular interactions through dipole-dipole interactions and $\pi$-$\pi$ stacking without considerably affecting the overall packing efficiency.  

    \begin{table}
      \centering
      \begin{tabular}{l c c}
        \hline
        \textbf{Polymer} & \textbf{$\Delta H_{vap}/H_{vap,PE}$ (\%)} & \textbf{$\Delta \rho/\rho_{_{PE}}$}\\
        \hline
        Aromatic Ester-linked PE & 21.0\% & 8.7\%\\
        Urethane-linked PE & 11.6\% & 4.4\%\\
        Amide-linked PE & 9.7\% & 3.01\% \\
        Oxalate-linked PE & 4.8\% & 3.4\% \\
        Ester-linked PE & 4.6\% & 3.2\% \\
        Urea-linked PE & 4.2\% & 1.9\% \\
        Carbonate-linked PE & 4.08\% & 2.7\% \\
        Siloxane-linked PE & 4.0\% & 2.6\% \\
        Anhydride-linked PE & 2.4\% & 2.5\% \\
        Acetal-linked PE & 0.8\% & 1.5\% \\
        \hline
      \end{tabular}
      \caption{Heat of vaporization and density increment relative to polyethylene at 373 K and a PE-backbone length of 80 united atoms.}
      \label{tab:heatvap}
    \end{table}

The modified PE with the next most significant increment is the urethane-linked PE with a value 11.6\% higher than that of PE. Next, at 373 K and for the longest backbone length, the amide-linked PE closely follows the trend of the urethane-liked PE with an increment of 9.7\%, almost 3 times higher than the increment in density. Both the urethane-linked PE and amide-linked PE share a very similar structure (Figure \ref{fig:structures}) with two functional groups separated by 6 methylene units in the center of its structure. Both urethanes and amides are able to form strong hydrogen bonds. The major difference lies in the additional oxygen in the urethane's chemical structure.

The oxalate, ester, urea, carbonate, and siloxane-linked PE show a very similar increment of around 4\% compared to polyethylene. Both the oxalate and the ester linkage contain two ester groups in their chemical structure. The ester linkage has two methylene units spacing their ester groups, whereas the oxalate linkage has no spacing between them. Even though the oxalate's rigid structure may facilitate better packing, the results show very insignificant differences between the heat of vaporization and the density of these two cleavable bonds. This implies that when the spacers between functionalities are short enough, the differences in the strength of intermolecular interactions and packing are negligible. Although the carbonate-linked PE would be expected to have lower values for heat of vaporization than an oxalate group or an ester-linkage, urea groups often form stronger hydrogen bond interactions due to the presence of two hydrogen bond donors on the nitrogen atoms. However, the results show that having two ester groups are seemingly equivalent to having an urea group in terms of the strength of intermolecular interactions. Surprisingly, the siloxane group has a relative increase in heat of vaporization comparable to that of the urea group. Since siloxanes do not generally form strong hydrogen bonds, this may seem counterintuitive. However, this could be explained by the polarizability of silicon atoms, which leads to an increase in London dispersion forces. The semi-polar Si-O bonds within a non-polar hydrocarbon could induce local dipoles and therefore increase the value for the heat of vaporization.

The anhydride-linked PE and the acetal-linked PE have the closest agreement with the unmodified PE. The anhydride, containing two carbonyl groups, enhances intermolecular forces through dipole-dipole interactions compared to the unmodified PE. For acetals, the presence of oxygen also introduces some lower dipole-dipole interactions. However, at longer chain lengths such contributions become insignificant.

\begin{figure}[ht]
    \centering
    \includegraphics[width=1.00\textwidth]{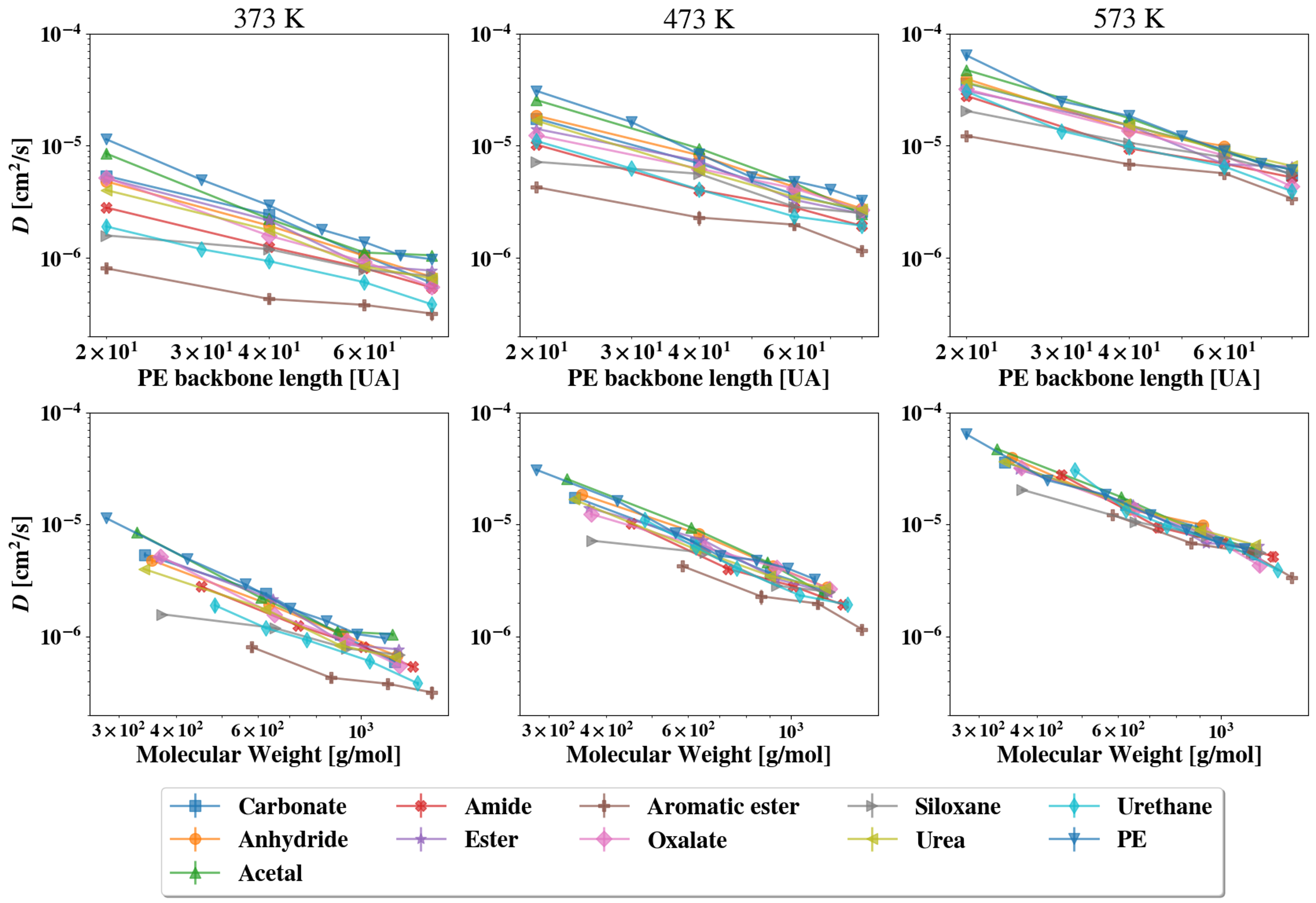}
    \caption{Self-diffusion coefficient for cleavable bond-modified polyethylene at 373, 473, and 573 K as a function of number of PE backbone united atoms and molecular weight.}
    \label{fig:diffusion_gromos}
\end{figure}

\subsection*{Diffusion}

 Figure \ref{fig:diffusion_gromos} shows diffusion coefficients for all cleavable bond-modified PEs at 373, 473, and 573 K, for PE backbone lengths up to 80 united atoms. At elevated temperatures, 573 K, for most of the PE-like melt, all curves collapse into a single line. At such temperatures, the segmental motion of the chains is less influenced by the specific nature of the modifying groups. The thermal energy is sufficiently high to overcome different intermolecular forces present in the modified PEs, leading to an increment of chain flexibility and similar diffusion values. As temperature decreases, the differences between diffusion coefficients of molecules with the same molecular weight become more important. 

At lower temperatures, the structural features of each of the cleavable bond-modified PEs affect the diffusion of the molecules. The chains with more hindrances to molecular motion due to stronger intermolecular forces, such as the aromatic ester-linked and urethane-linked PE, exhibit lower self-diffusion coefficients at lower temperatures. Conversely, polymers with more flexible structures like the unmodified linear PE and the acetal-linked PE exhibit relatively higher self-diffusion coefficients under the same conditions. Lower diffusion values and differences in slopes of diffusion as a function of molecular weight indicate that the chemical structure of the cleavable bonds does have a significant impact on the physical behavior of the polymer closer to the glass transition or crystallization temperatures.

\subsection*{Crystallization Temperature}

\begin{figure}[!ht]
    \centering
    \includegraphics[width=1.0\textwidth]{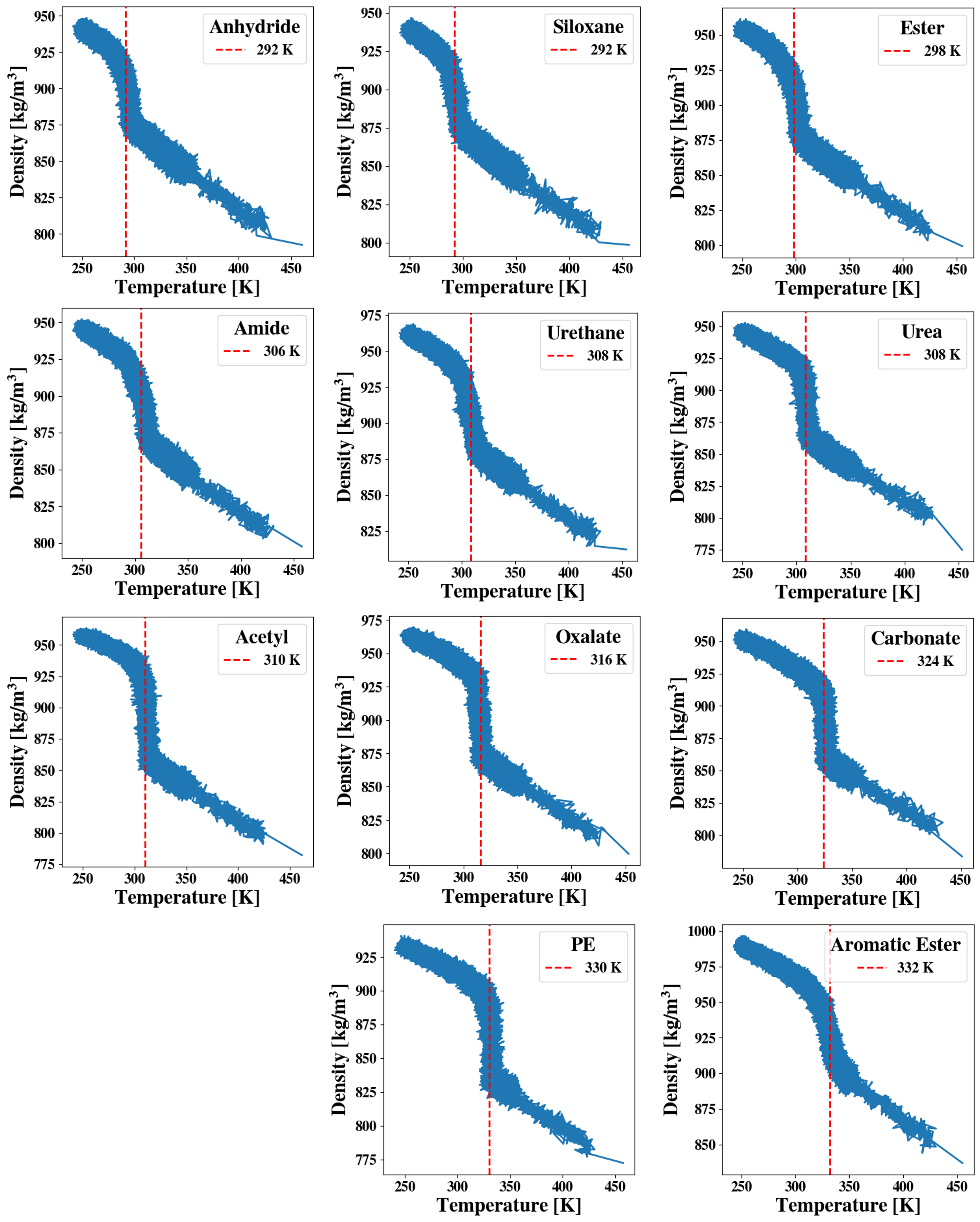}
    \caption{Crystallization temperatures for all cleavable bond-modified PE chains (blue) and unmodified PE (red) at a cooling rate of 0.05 K/ns.}
    \label{fig:density_vs_temp}
\end{figure}

\begin{figure}[!ht]
    \centering
    \includegraphics[width=0.9\textwidth]{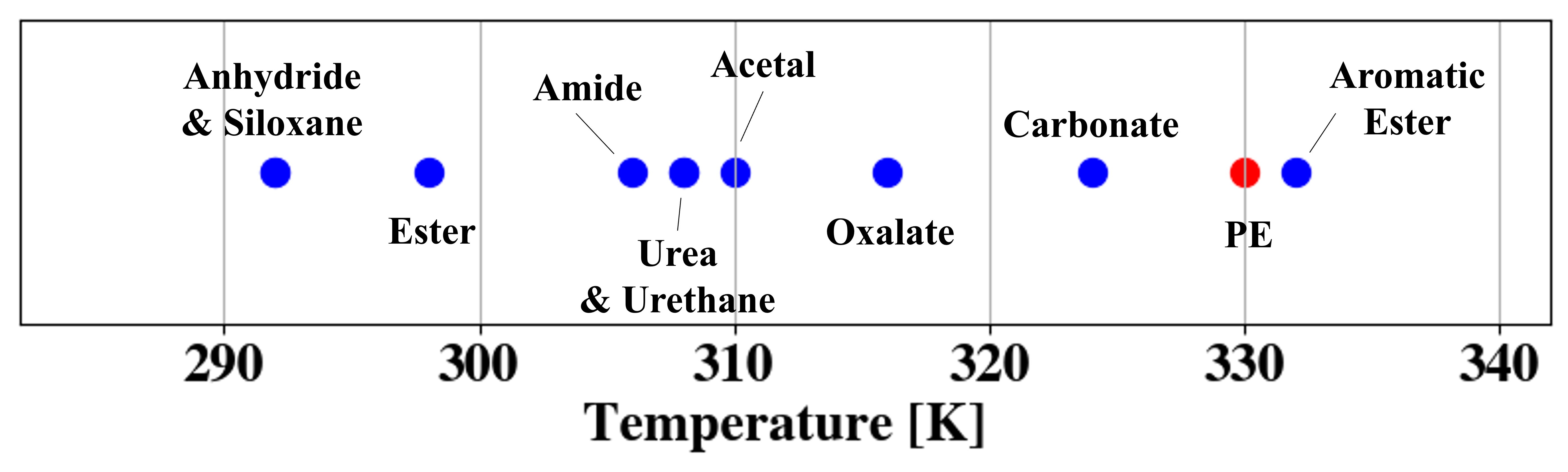}
    \caption{Crystallization temperatures for all cleavable bond-modified PE chains (blue) and unmodified PE (red) at a cooling rate of 0.05 K/ns.}
    \label{fig:t_crystal}
\end{figure}

Figure \ref{fig:density_vs_temp} shows the evolution of the density as a function of temperature for the different cleavable bond-modified polyethylenes, obtained through an annealing process. A sudden drop in density indicates a phase transition, in this case crystallization. A summary of the crystallization temperatures for a cooling rate of 0.05 K/ns is shown in Figure \ref{fig:t_crystal}. Note that the experimental cooling rates accessible to simulations are several orders of magnitude lower than those used in experiments. 

All polymers, except for the aromatic ester-linked PE, display a lower crystallization temperature than PE. The presence of the bulky aromatic ester groups allows the material to retain its semi-crystalline state over a wider temperature range. Since most of the hydrogen bonding linkages have lower crystallization temperatures, the higher value is likely explained by the strength and rigidity of $\pi$-$\pi$ stacking interactions imparted by the aromatic rings, which facilitate a more ordered packing and help nucleate crystallization early on.

\begin{figure}[!hb]
    \centering
    \includegraphics[width=1.0\textwidth]{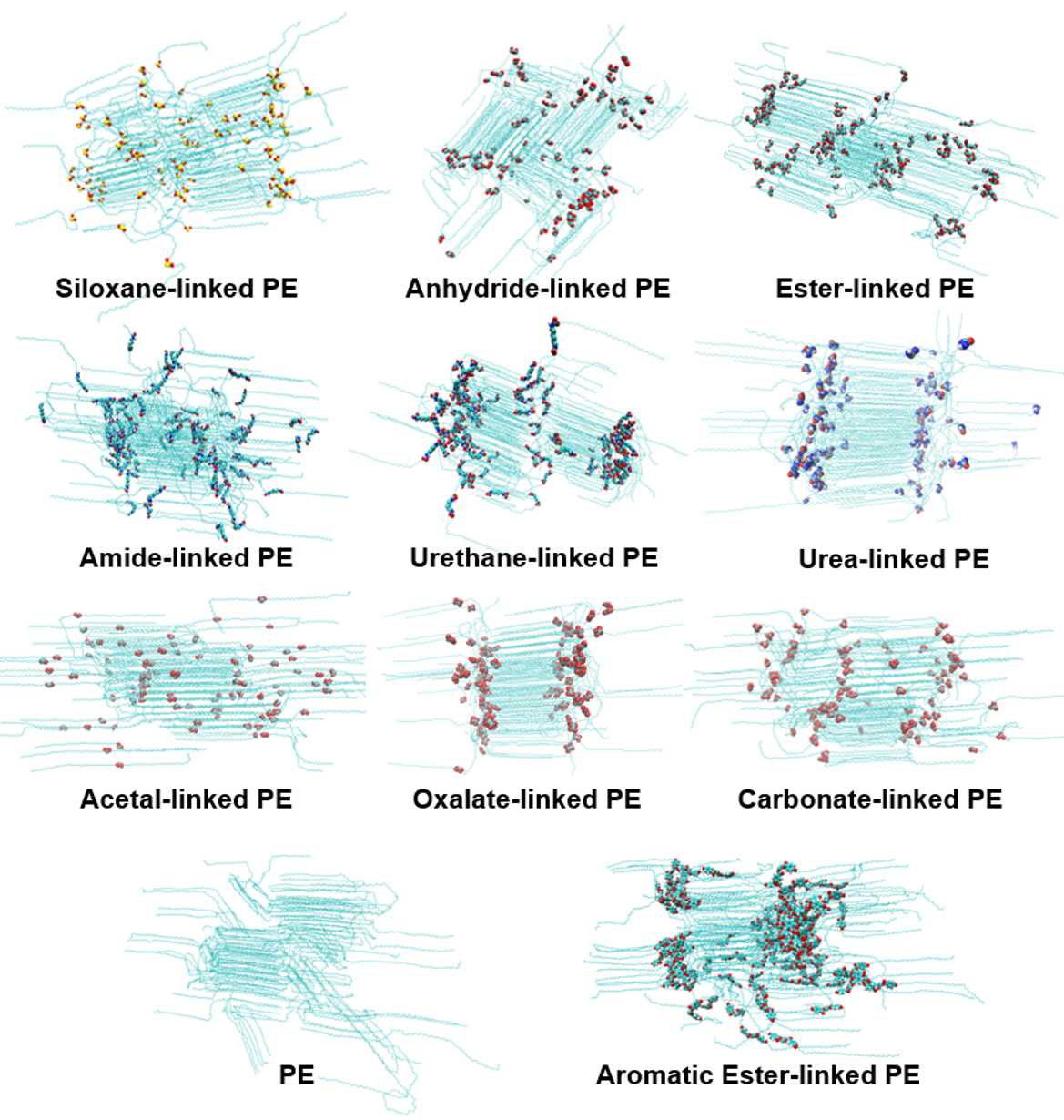}
    \caption{Cleavable bond-modified PE and PE structures at 150 K obtained through an annealing process at a rate of 0.05 K/ns.}
    \label{fig:crystal_snapshots}
\end{figure}

Figure \ref{fig:crystal_snapshots} shows snapshots of the semi-crystalline structures resulting from the annealing of PE ordered from lowest to highest crystallization temperature. From the annealed structures, we can observe that even the semi-polar siloxane linkages exhibit a tendency to aggregate in the resulting semi-crystalline structures. In a previous study, Menges \emph{et al}\cite{menges2007characterization} examined the crystalline structure of a long-chain aliphatic polyester featuring cleavable bonds with the same ester linkage considered here. The structures obtained at 150 K are in good qualitative agreement with the structures reported from X-ray Scattering and Nuclear Magnetic Resonance. The semi-crystalline behavior observed here supports the findings of these authors regarding intermolecular interactions: a few functional groups introduced into a purely aliphatic polymer are sufficient to control the morphology of polyethylene-like polymers.
\newpage

\newpage
\newpage
\section{Conclusions}
In this work, we compared the effect of different cleavable bonds on the structure of various PE chains. As expected, as the number of methylene units increase in the chain, all properties converge to those of polyethylene. A notable exception is provided by for the aromatic ester linkage, which exhibits a density that is considerably and consistently higher than that of unmodified PE. Such a higher density could be turned to an advantage in several applications of polyethylene. More generally, in all cases the observed differences in density are sufficiently important to lead to differences in processability, particularly as the molecular weight between cleavable bonds decreases. In this regard, the acetal linkages appear to exhibit the smallest disruption to the underlying PE backbone, whereas the aromatic ester linkage is the most disruptive. Similarly, at elevated temperatures the diffusion coefficients of all polymers are similar (except the aromatic ester linked materials), and are proportional to the molecular weight. At lower temperatures, the differences in diffusion coefficient become more pronounced. The aromatic ester-linked PE has the closest crystallization temperature to polyethylene and also the highest crystallization temperature of all structures studied in this work. For all the semi-crystalline structures considered here, we find that the cleavable bonds are excluded from the crystalline domains of the material, leading to the formation of aggregates within the material. We speculate, however, that in the case of aromatic-ester linkages the presense of $\pi$-$\pi$ stacking interactions helps nucleate the formation of ordered structures that facilitate crystallization.

\begin{acknowledgement}
This work was supported as part of the Center for Plastics Innovation, an Energy Frontier Research Center funded by the U.S. Department of Energy, Office of Science, Basic Energy Sciences at the University of Chicago under Grant Number DE-SC0021166. 
R.A. acknowledges support by the Dutch Research Council (NWO Rubicon 019.202EN.028).

\end{acknowledgement}




\bibliography{MyBib}

\providecommand{\latin}[1]{#1}
\makeatletter
\providecommand{\doi}
  {\begingroup\let\do\@makeother\dospecials
  \catcode`\{=1 \catcode`\}=2 \doi@aux}
\providecommand{\doi@aux}[1]{\endgroup\texttt{#1}}
\makeatother
\providecommand*\mcitethebibliography{\thebibliography}
\csname @ifundefined\endcsname{endmcitethebibliography}  {\let\endmcitethebibliography\endthebibliography}{}
\begin{mcitethebibliography}{36}
\providecommand*\natexlab[1]{#1}
\providecommand*\mciteSetBstSublistMode[1]{}
\providecommand*\mciteSetBstMaxWidthForm[2]{}
\providecommand*\mciteBstWouldAddEndPuncttrue
  {\def\EndOfBibitem{\unskip.}}
\providecommand*\mciteBstWouldAddEndPunctfalse
  {\let\EndOfBibitem\relax}
\providecommand*\mciteSetBstMidEndSepPunct[3]{}
\providecommand*\mciteSetBstSublistLabelBeginEnd[3]{}
\providecommand*\EndOfBibitem{}
\mciteSetBstSublistMode{f}
\mciteSetBstMaxWidthForm{subitem}{(\alph{mcitesubitemcount})}
\mciteSetBstSublistLabelBeginEnd
  {\mcitemaxwidthsubitemform\space}
  {\relax}
  {\relax}

\bibitem[Giacovelli(2018)]{giacovelli2018single}
Giacovelli,~C. \emph{Single-use plastics: a roadmap for sustainability}; International Environmental Technology Centre, 2018\relax
\mciteBstWouldAddEndPuncttrue
\mciteSetBstMidEndSepPunct{\mcitedefaultmidpunct}
{\mcitedefaultendpunct}{\mcitedefaultseppunct}\relax
\EndOfBibitem
\bibitem[Geyer \latin{et~al.}(2017)Geyer, Jambeck, and Law]{geyer2017production}
Geyer,~R.; Jambeck,~J.~R.; Law,~K.~L. Production, use, and fate of all plastics ever made. \emph{Science advances} \textbf{2017}, \emph{3}, e1700782\relax
\mciteBstWouldAddEndPuncttrue
\mciteSetBstMidEndSepPunct{\mcitedefaultmidpunct}
{\mcitedefaultendpunct}{\mcitedefaultseppunct}\relax
\EndOfBibitem
\bibitem[Jambeck \latin{et~al.}(2015)Jambeck, Geyer, Wilcox, Siegler, Perryman, Andrady, Narayan, and Law]{jambeck2015plastic}
Jambeck,~J.~R.; Geyer,~R.; Wilcox,~C.; Siegler,~T.~R.; Perryman,~M.; Andrady,~A.; Narayan,~R.; Law,~K.~L. Plastic waste inputs from land into the ocean. \emph{Science} \textbf{2015}, \emph{347}, 768--771\relax
\mciteBstWouldAddEndPuncttrue
\mciteSetBstMidEndSepPunct{\mcitedefaultmidpunct}
{\mcitedefaultendpunct}{\mcitedefaultseppunct}\relax
\EndOfBibitem
\bibitem[Ragaert \latin{et~al.}(2017)Ragaert, Delva, and Van~Geem]{ragaert2017mechanical}
Ragaert,~K.; Delva,~L.; Van~Geem,~K. Mechanical and chemical recycling of solid plastic waste. \emph{Waste management} \textbf{2017}, \emph{69}, 24--58\relax
\mciteBstWouldAddEndPuncttrue
\mciteSetBstMidEndSepPunct{\mcitedefaultmidpunct}
{\mcitedefaultendpunct}{\mcitedefaultseppunct}\relax
\EndOfBibitem
\bibitem[Vollmer \latin{et~al.}(2020)Vollmer, Jenks, Roelands, White, van Harmelen, de~Wild, van Der~Laan, Meirer, Keurentjes, and Weckhuysen]{vollmer2020beyond}
Vollmer,~I.; Jenks,~M.~J.; Roelands,~M.~C.; White,~R.~J.; van Harmelen,~T.; de~Wild,~P.; van Der~Laan,~G.~P.; Meirer,~F.; Keurentjes,~J.~T.; Weckhuysen,~B.~M. Beyond mechanical recycling: Giving new life to plastic waste. \emph{Angewandte Chemie International Edition} \textbf{2020}, \emph{59}, 15402--15423\relax
\mciteBstWouldAddEndPuncttrue
\mciteSetBstMidEndSepPunct{\mcitedefaultmidpunct}
{\mcitedefaultendpunct}{\mcitedefaultseppunct}\relax
\EndOfBibitem
\bibitem[Lopez \latin{et~al.}(2017)Lopez, Artetxe, Amutio, Bilbao, and Olazar]{lopez2017thermochemical}
Lopez,~G.; Artetxe,~M.; Amutio,~M.; Bilbao,~J.; Olazar,~M. Thermochemical routes for the valorization of waste polyolefinic plastics to produce fuels and chemicals. A review. \emph{Renewable and Sustainable Energy Reviews} \textbf{2017}, \emph{73}, 346--368\relax
\mciteBstWouldAddEndPuncttrue
\mciteSetBstMidEndSepPunct{\mcitedefaultmidpunct}
{\mcitedefaultendpunct}{\mcitedefaultseppunct}\relax
\EndOfBibitem
\bibitem[Epps~III \latin{et~al.}(2021)Epps~III, Korley, Yan, Beers, and Burt]{epps2021sustainability}
Epps~III,~T.~H.; Korley,~L.~T.; Yan,~T.; Beers,~K.~L.; Burt,~T.~M. Sustainability of synthetic plastics: Considerations in materials life-cycle management. \emph{JACS Au} \textbf{2021}, \emph{2}, 3--11\relax
\mciteBstWouldAddEndPuncttrue
\mciteSetBstMidEndSepPunct{\mcitedefaultmidpunct}
{\mcitedefaultendpunct}{\mcitedefaultseppunct}\relax
\EndOfBibitem
\bibitem[Hinton \latin{et~al.}(2022)Hinton, Talley, Kots, Le, Zhang, Mackay, Kunjapur, Bai, Vlachos, Watson, \latin{et~al.} others]{hinton2022innovations}
Hinton,~Z.~R.; Talley,~M.~R.; Kots,~P.~A.; Le,~A.~V.; Zhang,~T.; Mackay,~M.~E.; Kunjapur,~A.~M.; Bai,~P.; Vlachos,~D.~G.; Watson,~M.~P.; others Innovations toward the valorization of plastics waste. \emph{Annual Review of Materials Research} \textbf{2022}, \emph{52}, 249--280\relax
\mciteBstWouldAddEndPuncttrue
\mciteSetBstMidEndSepPunct{\mcitedefaultmidpunct}
{\mcitedefaultendpunct}{\mcitedefaultseppunct}\relax
\EndOfBibitem
\bibitem[Zhao \latin{et~al.}(2022)Zhao, Korey, Li, Copenhaver, Tekinalp, Celik, Kalaitzidou, Ruan, Ragauskas, and Ozcan]{zhao2022plastic}
Zhao,~X.; Korey,~M.; Li,~K.; Copenhaver,~K.; Tekinalp,~H.; Celik,~S.; Kalaitzidou,~K.; Ruan,~R.; Ragauskas,~A.~J.; Ozcan,~S. Plastic waste upcycling toward a circular economy. \emph{Chemical Engineering Journal} \textbf{2022}, \emph{428}, 131928\relax
\mciteBstWouldAddEndPuncttrue
\mciteSetBstMidEndSepPunct{\mcitedefaultmidpunct}
{\mcitedefaultendpunct}{\mcitedefaultseppunct}\relax
\EndOfBibitem
\bibitem[Coates and Getzler(2020)Coates, and Getzler]{coates2020chemical}
Coates,~G.~W.; Getzler,~Y.~D. Chemical recycling to monomer for an ideal, circular polymer economy. \emph{Nature Reviews Materials} \textbf{2020}, \emph{5}, 501--516\relax
\mciteBstWouldAddEndPuncttrue
\mciteSetBstMidEndSepPunct{\mcitedefaultmidpunct}
{\mcitedefaultendpunct}{\mcitedefaultseppunct}\relax
\EndOfBibitem
\bibitem[Tournier \latin{et~al.}(2020)Tournier, Topham, Gilles, David, Folgoas, Moya-Leclair, Kamionka, Desrousseaux, Texier, Gavalda, \latin{et~al.} others]{tournier2020engineered}
Tournier,~V.; Topham,~C.; Gilles,~A.; David,~B.; Folgoas,~C.; Moya-Leclair,~E.; Kamionka,~E.; Desrousseaux,~M.-L.; Texier,~H.; Gavalda,~S.; others An engineered PET depolymerase to break down and recycle plastic bottles. \emph{Nature} \textbf{2020}, \emph{580}, 216--219\relax
\mciteBstWouldAddEndPuncttrue
\mciteSetBstMidEndSepPunct{\mcitedefaultmidpunct}
{\mcitedefaultendpunct}{\mcitedefaultseppunct}\relax
\EndOfBibitem
\bibitem[Stempfle \latin{et~al.}(2016)Stempfle, Ortmann, and Mecking]{stempfle2016long}
Stempfle,~F.; Ortmann,~P.; Mecking,~S. Long-chain aliphatic polymers to bridge the gap between semicrystalline polyolefins and traditional polycondensates. \emph{Chemical reviews} \textbf{2016}, \emph{116}, 4597--4641\relax
\mciteBstWouldAddEndPuncttrue
\mciteSetBstMidEndSepPunct{\mcitedefaultmidpunct}
{\mcitedefaultendpunct}{\mcitedefaultseppunct}\relax
\EndOfBibitem
\bibitem[H{\"a}u{\ss}ler \latin{et~al.}(2021)H{\"a}u{\ss}ler, Eck, Rothauer, and Mecking]{haussler2021closed}
H{\"a}u{\ss}ler,~M.; Eck,~M.; Rothauer,~D.; Mecking,~S. Closed-loop recycling of polyethylene-like materials. \emph{Nature} \textbf{2021}, \emph{590}, 423--427\relax
\mciteBstWouldAddEndPuncttrue
\mciteSetBstMidEndSepPunct{\mcitedefaultmidpunct}
{\mcitedefaultendpunct}{\mcitedefaultseppunct}\relax
\EndOfBibitem
\bibitem[Harmandaris \latin{et~al.}(2003)Harmandaris, Mavrantzas, Theodorou, Kr{\"o}ger, Ramirez, {\"O}ttinger, and Vlassopoulos]{harmandaris2003crossover}
Harmandaris,~V.; Mavrantzas,~V.; Theodorou,~D.; Kr{\"o}ger,~M.; Ramirez,~J.; {\"O}ttinger,~H.~C.; Vlassopoulos,~D. Crossover from the rouse to the entangled polymer melt regime: signals from long, detailed atomistic molecular dynamics simulations, supported by rheological experiments. \emph{Macromolecules} \textbf{2003}, \emph{36}, 1376--1387\relax
\mciteBstWouldAddEndPuncttrue
\mciteSetBstMidEndSepPunct{\mcitedefaultmidpunct}
{\mcitedefaultendpunct}{\mcitedefaultseppunct}\relax
\EndOfBibitem
\bibitem[Mondello \latin{et~al.}(1998)Mondello, Grest, Webb~III, and Peczak]{mondello1998dynamics}
Mondello,~M.; Grest,~G.~S.; Webb~III,~E.~B.; Peczak,~P. Dynamics of n-alkanes: Comparison to Rouse model. \emph{The Journal of chemical physics} \textbf{1998}, \emph{109}, 798--805\relax
\mciteBstWouldAddEndPuncttrue
\mciteSetBstMidEndSepPunct{\mcitedefaultmidpunct}
{\mcitedefaultendpunct}{\mcitedefaultseppunct}\relax
\EndOfBibitem
\bibitem[Ramos \latin{et~al.}(2008)Ramos, Vega, Theodorou, and Martinez-Salazar]{ramos2008entanglement}
Ramos,~J.; Vega,~J.~F.; Theodorou,~D.~N.; Martinez-Salazar,~J. Entanglement relaxation time in polyethylene: Simulation versus experimental data. \emph{Macromolecules} \textbf{2008}, \emph{41}, 2959--2962\relax
\mciteBstWouldAddEndPuncttrue
\mciteSetBstMidEndSepPunct{\mcitedefaultmidpunct}
{\mcitedefaultendpunct}{\mcitedefaultseppunct}\relax
\EndOfBibitem
\bibitem[Ramos \latin{et~al.}(2018)Ramos, Vega, and Mart{\'\i}nez-Salazar]{ramos2018predicting}
Ramos,~J.; Vega,~J.~F.; Mart{\'\i}nez-Salazar,~J. Predicting experimental results for polyethylene by computer simulation. \emph{European Polymer Journal} \textbf{2018}, \emph{99}, 298--331\relax
\mciteBstWouldAddEndPuncttrue
\mciteSetBstMidEndSepPunct{\mcitedefaultmidpunct}
{\mcitedefaultendpunct}{\mcitedefaultseppunct}\relax
\EndOfBibitem
\bibitem[da~Silva \latin{et~al.}(2020)da~Silva, Silva, Tavares, Fleming, and Horta]{da2020all}
da~Silva,~G.~C.; Silva,~G.~M.; Tavares,~F.~W.; Fleming,~F.~P.; Horta,~B.~A. Are all-atom any better than united-atom force fields for the description of liquid properties of alkanes? \emph{Journal of Molecular Modeling} \textbf{2020}, \emph{26}, 1--17\relax
\mciteBstWouldAddEndPuncttrue
\mciteSetBstMidEndSepPunct{\mcitedefaultmidpunct}
{\mcitedefaultendpunct}{\mcitedefaultseppunct}\relax
\EndOfBibitem
\bibitem[da~Silva \latin{et~al.}(2022)da~Silva, Silva, Tavares, Fleming, and Horta]{da2022all}
da~Silva,~G.~C.; Silva,~G.~M.; Tavares,~F.~W.; Fleming,~F.~P.; Horta,~B.~A. Are all-atom any better than united-atom force fields for the description of liquid properties of alkanes? 2. A systematic study considering different chain lengths. \emph{Journal of Molecular Liquids} \textbf{2022}, \emph{354}, 118829\relax
\mciteBstWouldAddEndPuncttrue
\mciteSetBstMidEndSepPunct{\mcitedefaultmidpunct}
{\mcitedefaultendpunct}{\mcitedefaultseppunct}\relax
\EndOfBibitem
\bibitem[Schuler \latin{et~al.}(2001)Schuler, Daura, and Van~Gunsteren]{schuler2001improved}
Schuler,~L.~D.; Daura,~X.; Van~Gunsteren,~W.~F. An improved GROMOS96 force field for aliphatic hydrocarbons in the condensed phase. \emph{Journal of computational chemistry} \textbf{2001}, \emph{22}, 1205--1218\relax
\mciteBstWouldAddEndPuncttrue
\mciteSetBstMidEndSepPunct{\mcitedefaultmidpunct}
{\mcitedefaultendpunct}{\mcitedefaultseppunct}\relax
\EndOfBibitem
\bibitem[Dewar and Thiel(1977)Dewar, and Thiel]{dewar1977ground}
Dewar,~M.~J.; Thiel,~W. Ground states of molecules. 38. The MNDO method. Approximations and parameters. \emph{Journal of the American Chemical Society} \textbf{1977}, \emph{99}, 4899--4907\relax
\mciteBstWouldAddEndPuncttrue
\mciteSetBstMidEndSepPunct{\mcitedefaultmidpunct}
{\mcitedefaultendpunct}{\mcitedefaultseppunct}\relax
\EndOfBibitem
\bibitem[Stewart(2013)]{stewart2013optimization}
Stewart,~J.~J. Optimization of parameters for semiempirical methods VI: more modifications to the NDDO approximations and re-optimization of parameters. \emph{Journal of molecular modeling} \textbf{2013}, \emph{19}, 1--32\relax
\mciteBstWouldAddEndPuncttrue
\mciteSetBstMidEndSepPunct{\mcitedefaultmidpunct}
{\mcitedefaultendpunct}{\mcitedefaultseppunct}\relax
\EndOfBibitem
\bibitem[Stroet \latin{et~al.}(2018)Stroet, Caron, Visscher, Geerke, Malde, and Mark]{stroet2018automated}
Stroet,~M.; Caron,~B.; Visscher,~K.~M.; Geerke,~D.~P.; Malde,~A.~K.; Mark,~A.~E. Automated topology builder version 3.0: Prediction of solvation free enthalpies in water and hexane. \emph{Journal of chemical theory and computation} \textbf{2018}, \emph{14}, 5834--5845\relax
\mciteBstWouldAddEndPuncttrue
\mciteSetBstMidEndSepPunct{\mcitedefaultmidpunct}
{\mcitedefaultendpunct}{\mcitedefaultseppunct}\relax
\EndOfBibitem
\bibitem[Gr{\"u}newald \latin{et~al.}(2022)Gr{\"u}newald, Alessandri, Kroon, Monticelli, Souza, and Marrink]{grunewald2022polyply}
Gr{\"u}newald,~F.; Alessandri,~R.; Kroon,~P.~C.; Monticelli,~L.; Souza,~P.~C.; Marrink,~S.~J. Polyply; a python suite for facilitating simulations of macromolecules and nanomaterials. \emph{Nature Communications} \textbf{2022}, \emph{13}, 68\relax
\mciteBstWouldAddEndPuncttrue
\mciteSetBstMidEndSepPunct{\mcitedefaultmidpunct}
{\mcitedefaultendpunct}{\mcitedefaultseppunct}\relax
\EndOfBibitem
\bibitem[Van Der~Spoel \latin{et~al.}(2005)Van Der~Spoel, Lindahl, Hess, Groenhof, Mark, and Berendsen]{van2005gromacs}
Van Der~Spoel,~D.; Lindahl,~E.; Hess,~B.; Groenhof,~G.; Mark,~A.~E.; Berendsen,~H.~J. GROMACS: fast, flexible, and free. \emph{Journal of computational chemistry} \textbf{2005}, \emph{26}, 1701--1718\relax
\mciteBstWouldAddEndPuncttrue
\mciteSetBstMidEndSepPunct{\mcitedefaultmidpunct}
{\mcitedefaultendpunct}{\mcitedefaultseppunct}\relax
\EndOfBibitem
\bibitem[Pronk \latin{et~al.}(2013)Pronk, P{\'a}ll, Schulz, Larsson, Bjelkmar, Apostolov, Shirts, Smith, Kasson, Van Der~Spoel, \latin{et~al.} others]{pronk2013gromacs}
Pronk,~S.; P{\'a}ll,~S.; Schulz,~R.; Larsson,~P.; Bjelkmar,~P.; Apostolov,~R.; Shirts,~M.~R.; Smith,~J.~C.; Kasson,~P.~M.; Van Der~Spoel,~D.; others GROMACS 4.5: a high-throughput and highly parallel open source molecular simulation toolkit. \emph{Bioinformatics} \textbf{2013}, \emph{29}, 845--854\relax
\mciteBstWouldAddEndPuncttrue
\mciteSetBstMidEndSepPunct{\mcitedefaultmidpunct}
{\mcitedefaultendpunct}{\mcitedefaultseppunct}\relax
\EndOfBibitem
\bibitem[Berendsen \latin{et~al.}(1995)Berendsen, van~der Spoel, and van Drunen]{berendsen1995gromacs}
Berendsen,~H.~J.; van~der Spoel,~D.; van Drunen,~R. GROMACS: A message-passing parallel molecular dynamics implementation. \emph{Computer physics communications} \textbf{1995}, \emph{91}, 43--56\relax
\mciteBstWouldAddEndPuncttrue
\mciteSetBstMidEndSepPunct{\mcitedefaultmidpunct}
{\mcitedefaultendpunct}{\mcitedefaultseppunct}\relax
\EndOfBibitem
\bibitem[Abraham \latin{et~al.}(2015)Abraham, Murtola, Schulz, P{\'a}ll, Smith, Hess, and Lindahl]{abraham2015gromacs}
Abraham,~M.~J.; Murtola,~T.; Schulz,~R.; P{\'a}ll,~S.; Smith,~J.~C.; Hess,~B.; Lindahl,~E. GROMACS: High performance molecular simulations through multi-level parallelism from laptops to supercomputers. \emph{SoftwareX} \textbf{2015}, \emph{1}, 19--25\relax
\mciteBstWouldAddEndPuncttrue
\mciteSetBstMidEndSepPunct{\mcitedefaultmidpunct}
{\mcitedefaultendpunct}{\mcitedefaultseppunct}\relax
\EndOfBibitem
\bibitem[Maginn \latin{et~al.}(2019)Maginn, Messerly, Carlson, Roe, and Elliot]{maginn2019best}
Maginn,~E.~J.; Messerly,~R.~A.; Carlson,~D.~J.; Roe,~D.~R.; Elliot,~J.~R. Best practices for computing transport properties 1. Self-diffusivity and viscosity from equilibrium molecular dynamics [article v1. 0]. \emph{Living Journal of Computational Molecular Science} \textbf{2019}, \emph{1}, 6324--6324\relax
\mciteBstWouldAddEndPuncttrue
\mciteSetBstMidEndSepPunct{\mcitedefaultmidpunct}
{\mcitedefaultendpunct}{\mcitedefaultseppunct}\relax
\EndOfBibitem
\bibitem[Basconi and Shirts(2013)Basconi, and Shirts]{basconi2013effects}
Basconi,~J.~E.; Shirts,~M.~R. Effects of temperature control algorithms on transport properties and kinetics in molecular dynamics simulations. \emph{Journal of chemical theory and computation} \textbf{2013}, \emph{9}, 2887--2899\relax
\mciteBstWouldAddEndPuncttrue
\mciteSetBstMidEndSepPunct{\mcitedefaultmidpunct}
{\mcitedefaultendpunct}{\mcitedefaultseppunct}\relax
\EndOfBibitem
\bibitem[Bussi \latin{et~al.}(2007)Bussi, Donadio, and Parrinello]{bussi2007canonical}
Bussi,~G.; Donadio,~D.; Parrinello,~M. Canonical sampling through velocity rescaling. \emph{The Journal of chemical physics} \textbf{2007}, \emph{126}\relax
\mciteBstWouldAddEndPuncttrue
\mciteSetBstMidEndSepPunct{\mcitedefaultmidpunct}
{\mcitedefaultendpunct}{\mcitedefaultseppunct}\relax
\EndOfBibitem
\bibitem[Nos{\'e} and Klein(1983)Nos{\'e}, and Klein]{nose1983constant}
Nos{\'e},~S.; Klein,~M. Constant pressure molecular dynamics for molecular systems. \emph{Molecular Physics} \textbf{1983}, \emph{50}, 1055--1076\relax
\mciteBstWouldAddEndPuncttrue
\mciteSetBstMidEndSepPunct{\mcitedefaultmidpunct}
{\mcitedefaultendpunct}{\mcitedefaultseppunct}\relax
\EndOfBibitem
\bibitem[Parrinello and Rahman(1981)Parrinello, and Rahman]{parrinello1981polymorphic}
Parrinello,~M.; Rahman,~A. Polymorphic transitions in single crystals: A new molecular dynamics method. \emph{Journal of Applied physics} \textbf{1981}, \emph{52}, 7182--7190\relax
\mciteBstWouldAddEndPuncttrue
\mciteSetBstMidEndSepPunct{\mcitedefaultmidpunct}
{\mcitedefaultendpunct}{\mcitedefaultseppunct}\relax
\EndOfBibitem
\bibitem[Sun and Rigby(1997)Sun, and Rigby]{sun1997polysiloxanes}
Sun,~H.; Rigby,~D. Polysiloxanes: ab initio force field and structural, conformational and thermophysical properties. \emph{Spectrochimica Acta Part A: Molecular and Biomolecular Spectroscopy} \textbf{1997}, \emph{53}, 1301--1323\relax
\mciteBstWouldAddEndPuncttrue
\mciteSetBstMidEndSepPunct{\mcitedefaultmidpunct}
{\mcitedefaultendpunct}{\mcitedefaultseppunct}\relax
\EndOfBibitem
\bibitem[Menges \latin{et~al.}(2007)Menges, Penelle, Le~Fevere~de Ten~Hove, Jonas, and Schmidt-Rohr]{menges2007characterization}
Menges,~M.; Penelle,~J.; Le~Fevere~de Ten~Hove,~C.; Jonas,~A.~M.; Schmidt-Rohr,~K. Characterization of long-chain aliphatic polyesters: Crystalline and supramolecular structure of PE22, 4 elucidated by X-ray scattering and nuclear magnetic resonance. \emph{Macromolecules} \textbf{2007}, \emph{40}, 8714--8725\relax
\mciteBstWouldAddEndPuncttrue
\mciteSetBstMidEndSepPunct{\mcitedefaultmidpunct}
{\mcitedefaultendpunct}{\mcitedefaultseppunct}\relax
\EndOfBibitem
\end{mcitethebibliography}

\end{document}